\documentclass[debug]{rmaa}

\usepackage{hyperref}
\usepackage{rotating}
\usepackage{ulem}  

\usepackage{paralist}
\usepackage{tabularx,adjustbox,booktabs}

\usepackage{psfrag,color,ulem,gensymb}
\definecolor{deepblue}{rgb}{0,0,0.5}
\definecolor{deepred}{rgb}{0.6,0,0}
\definecolor{deepgreen}{rgb}{0,0.5,0}
\definecolor{darkgreen}{rgb}{0,0.6,0}

\usepackage[latin1]{inputenc}

\usepackage{textcomp}
\usepackage{xspace}
\usepackage{booktabs,caption}
\usepackage[flushleft]{threeparttable}

\newcommand{\ramoiii}{\hbox{$\xi_{\rm OIII}^{S_{0}}$}}

\newcommand{\ram}{\hbox{$\xi$}}

\newcommand{\gam}{\hbox{$\gamma$}}

\newcommand{\lam}{\hbox{$\lambda$}}

\newcommand{\nout}{\hbox{\it n$_{H}^{\rm o}$}}

\newcommand{\uout}{\hbox{\it {\rm U}$_{\rm o}$}}

\newcommand{\Tcut}{\hbox{\it T$_{cut}$}}
\newcommand{\Tbb}{\hbox{\it T$_{BB}$}}

\newcommand{\rpk}{\hbox{$r_{21}^{pk2}$}}

\newcommand{\Fpko}{\hbox{$F_{La12}^{pk1}$}}
\newcommand{\Fpkt}{\hbox{$F_{*}^{pk2}$}}

\newcommand{\cms}{\hbox{${\rm cm^{-2}}$}} 
\newcommand{\cmc}{\hbox{${\rm cm^{-3}}$}}

\newcommand{\ev}{\hbox{\sl eV}}

\newcommand{\alnu}{\hbox{$\alpha_{\nu}$}}
\newcommand{\aox}{\hbox{$\alpha_{OX}$}}

\newcommand{\auv}{\hbox{$\alpha_{UV}$}}
\newcommand{\afuv}{\hbox{$\alpha_{FUV}$}}
\newcommand{\abunoh}{\hbox{$12 + \log({\rm O/H}$}}

\newcommand{\ztot}{\hbox{$Z_{\rm tot}$}}

\newcommand{\zsol}{\hbox{$Z_{\sun}$}}

\newcommand{\Laasto}{\hbox{La$^{1}_{\star}$}}
\newcommand{\Laastd}{\hbox{La$^{2}_{\star}$}}

\newcommand{\kev}{\hbox{\sl keV}}

\newcommand{\degk}{{\degree}K}

\newcommand{\hi}{\hbox{H\,{\sc i}}}
\newcommand{\hii}{\hbox{H\,{\sc ii}}}
\newcommand{\hen}{\hbox{He$^{\rm 0}$}}
\newcommand{\hep}{\hbox{He$^{\rm +}$}}

\newcommand{\npp}{\hbox{N$^{\rm +2}$}}
\newcommand{\opp}{\hbox{O$^{\rm +2}$}}
\newcommand{\oppp}{\hbox{O$^{\rm +3}$}}
\newcommand{\nso}{\hbox{$^{\rm 1}{S}_{\rm 0}$}}
\newcommand{\ndd}{\hbox{$^{\rm 1}{D}_{\rm 2}$}}
\newcommand{\ddx}{\hbox{$^{\rm 2}{D}$}}
\newcommand{\dpx}{\hbox{$^{\rm 2}{P}$}}

\newcommand{\spp}{\hbox{S$^{\rm +2}$}}
\newcommand{\sppp}{\hbox{S$^{\rm +3}$}}

\newcommand{\hb}{\hbox{H$\beta$}}

\newcommand{\heii}{\hbox{He\,{\sc ii}}}

\newcommand{\oiiir}{\hbox{\lam4363\AA/\lam5007\AA}}

\newcommand{\oiiirb}{\hbox{\lam5007\AA/\lam4861\AA}}
\newcommand{\heiirb}{\hbox{\lam4686\AA/\lam4861\AA}}

\newcommand{\nii}{\hbox{[N\,{\sc ii}]}}

\newcommand{\ciii}{\hbox{C\,{\sc iii}]}}
\newcommand{\civ}{\hbox{C\,{\sc iv}}}

\newcommand{\ariv}{\hbox{[Ar\,{\sc iv}]}}

\newcommand{\Roiii}{\hbox{$R_{\rm OIII}$}}

\newcommand{\neiii}{\hbox{[Ne\,{\sc iii}]}}

\newcommand{\oiiix}{\hbox{O\,{\sc iii}}}
\newcommand{\oiii}{\hbox{[O\,{\sc iii}]}}

\newcommand{\oiiitw}{\hbox{[O\,{\sc iii}]\,\lam4363\AA}}

\newcommand{\oii}{\hbox{[O\,{\sc ii}]}}

\def\arcsec{\hbox{$^{\hbox{\rlap{\hbox{\lower4pt\hbox{$\,\prime\prime$}}
          }\hbox{$\frown$}}}$}}
\def\arcmin{\hbox{$^{\hbox{\rlap{\hbox{\lower4pt\hbox{$\;\prime$}}
          }\hbox{$\frown$}}}$}}

\newcommand{\out}[1]{}  

\newcommand{\map}{\hbox{{\sc mappings i}g}}
\newcommand{\ifla}{\hbox{{\sc i}g}}
\newcommand{\MAP}{\hbox{{\sc mappings i}}}
\newcommand{\OSALD}{{\sc osald}}

\usepackage{lscape}
\usepackage{listings}
\DeclareFixedFont{\ttb}{T1}{txtt}{bx}{n}{10} 
\DeclareFixedFont{\ttm}{T1}{txtt}{m}{n}{10}  

\title{Optimized Spectral energy distribution for Seyfert galaxies}

\author{
  Luc Binette\altaffilmark{1,2},
  Yair Krongold\altaffilmark{1},
  Sinhue A.R. Haro-Corzo\altaffilmark{3},
  Andrew Humphrey\altaffilmark{4,5},
  Sandy G. Morais\altaffilmark{4,6}
  }

\altaffiltext{1}{Instituto de Astronom\'\i{}a, UNAM, M\'exico}
\altaffiltext{2}{D\'epartement de physique, de g\'enie physique et d'optique,  Universit\'e Laval, Qu\'ebec}
\altaffiltext{3}{Escuela Nacional de Estudios Superiores (ENES) Unidad Morelia, UNAM, M\'exico}
\altaffiltext{4}{Instituto de Astrof\'{i}sica e Ci\^encias do Espa\c{c}o, Universidade do Porto, Portugal}
\altaffiltext{5}{DTx -- Digital Transformation CoLAB, University of Minho, Portugal}
\altaffiltext{6}{Departamento de F\'\i{}sica e Astronomia, Faculdade de Ci\^encias, Universidade do Porto, Portugal}

\shortauthor{Binette, Krongold, et\,al.}

\shorttitle{Active Nuclei SED}

\fulladdresses{
\item L. Binette and Y. Krongold: Instituto de Astronom\'\i{}a, Universidad Nacional Aut\'onoma de M\'exico, A.P. 70-264, 04510 M\'exico, D.F., M\'exico, Apartado Postal 70-264, Ciudad de M\'exico, CDMX, C.P. 04510, M\'exico. 
\item S.A.R Haro-Corzo: Escuela Nacional de Estudios Superiores (ENES) Unidad Morelia, UNAM, Antigua Carretera a Patzcuaro No. 8701 Col. Ex Hacienda de San Jos\'e de la Huerta C.P. 58190 Morelia, Michoac\'an, M\'exico.
\item A. Humphrey  and S.G. Morais: Instituto de Astrof\'\i{}sica e Ci\^encias do Espa\c{c}o, Universidade do Porto, CAUP, Rua das Estrelas, 4150-762, Porto, Portugal
\item A. Humphrey: DTx -- Digital Transformation CoLAB, Building 1, Azur\'em Campus, University of Minho, 4800-058 Guimar\~aes, Portugal
\item S.G. Morais: Departamento de F\'\i{}sica e Astronomia, Faculdade de Ci\^encias, Universidade do Porto, Rua do Campo Alegre 687, PT4169-007 Porto, Portugal

  }

\listofauthors{L. Binette, }
\indexauthor{Binette, L.}
\indexauthor{Krongold, Y.}
\indexauthor{Haro-Corzo, S.}
\indexauthor{Humphrey, A.}
\indexauthor{Morais, S.G.}
\abstract{The temperature predicted by photoionization models for the Narrow Line Region of Seyfert\,2  galaxies is lower than the value inferred from the observed \oiii\ \oiiir\ line ratio. We explore the possibility of considering a harder ionizing continuum than typically assumed. The spectral ionizing energy distribution, which can generate the observed \oiiir\ ratio, is characterized by a secondary continuum peak at 200\,\ev.  
}

\resumen{La temperatura predicha por modelos de fotoionizaci\'on de la regi\'on de l\'{\i}neas angostas (NLR) es inferior al valor deducido por el cociente de l\'{\i}neas \oiii\ \oiiir\ que se observa en galaxias Seyfert\,2. Exploramos la posibilidad de un continuo ionizante mucho m\'as duro que \'el que tipicamente se usa. La distribuci\'on de energ\'{\i}a espectral ionizante que logra reproducir el cociente de \oiiir\ observado se caracteriza por un segundo pico en el continuo a 200\,\ev. }

\addkeyword{galaxies: Seyfert}
\addkeyword{plasmas: }
\addkeyword{quasars: emission lines}
\addkeyword{accretion: accretion disks}

\begin{document}
\maketitle


\section{Introduction}
\label{sec:intro}

It has been proposed early on that photoionization is the excitation mechanism of the plasma associated to the Narrow Line Region of Active Galactic Nuclei (AGN) \citep[][and references therein]{OST78}. Prevailing photoionization models of the  Narrow-Line Region (NLR) of AGN consider a distribution of clouds that extends over a wide range of cloud densities and ionization parameter values, whether the targets are Type\,I \citep{Ba95,Kor97a} such as quasars, Seyfert\,1's and Broad-Line Radio Galaxies, or Type\,II objects \citep{Fg97,Ri14} which consist of Seyfert\,2's, QSO\,2's and Narrow-Line Radio Galaxies (NLRG). One difficulty reported by \citet{SB96}, \citet{Be06b}, \citet{VM08} and \citet{Dr15,Dr20b} is that the temperature predicted by photoionization models is lower than the value inferred from the observed \oiii\ \oiiir\ line ratio (hereafter labeled \Roiii). {This discrepancy defines the so-called ``temperature-problem'',  which is mentioned below and refers only to the \textit{spatially unresolved} NLR. Our basic assumption is that photoionization is the dominant excitation mechanism. We cannot rule out the presence of shocks, but combining shocks and photoionization in order to fit a sample of objects that share a similar temperature would a require  fine-tuning of both heating mechanism, which would not be a convincing procedure. We recognize that observations of the spatially out-flowing plasma, which are labelled Extended Narrow Line Region (ENLR), indicate in some Seyferts the presence of a much hotter plasma. For instance, the IFU MUSE/VLT observations of the Seyfert\,2 Circinus by \citet{FF21} reveal temperatures as high as 20\,000\,\degk\ within the ENLR, which standard photoionisation models cannot reproduce. The current work addresses only the NLR where we will assume that the dominant heating mechanism is photoionization, although we do rule out that other factors might affect the \Roiii\ ratio, such as a non-maxwellian electron energy distribution \citep{Mo21}, or the contribution of matter-bounded clouds to the emission line spectrum \citep{BWRS}.
}

The \Roiii\ ratio is a valid temperature diagnostic when the low density regime (LDR) applies (such as in \hii\ regions), that is, for plasma of densities $\la 10^4$\,\cmc, otherwise collisional deexcitation becomes important, which causes the \Roiii\ ratio to increase with density, independently of the temperature. The presence of significant collisional deexcitation appears to be the norm in Type\,I AGN, as shown by the work of \citet[hereafter BL05]{BL05} who compared the \Roiii\ values observed in 30 quasars\footnote{The term `quasar' is used to refer to Type\,I AGN.}. In Type\,II AGN, however, there is indirect evidence that collisional deexcitation is not dominant. For instance, the \Roiii\ ratios of Seyfert\,2's are found to be similar to those observed within the spatially resolved component, the so-called Extended Narrow Line Region (ENLR), where LDR conditions are known to apply. More direct evidence of LDR conditions in Type\,II nuclei was recently presented by \citet[hereafter BVM]{BVM} who used the measurements  of the \ariv\ $\lambda\lambda$4711,40\AA\ doublet ratio observed by \citet[hereafter Kos78]{Kos78} in seven Seyfert\,2's, which indicated that the densities were $\la 10^4$\,\cmc. The average NLR temperature inferred was 13\,500\,\degk, which standard single-zone photoionization models cannot reproduce. 

In the present work, we will investigate  whether an optimization of the spectral energy distribution (SED) of the ionizing source might contribute to the resolution of the temperature problem. Since we do not directly observe the  far-UV region of the ionizing continuum due to interstellar absorption, the current study is speculative in nature. Alternative  interpretations of the temperature discrepancy will be the subject of future publications.


\begin{figure}[!t]
  \includegraphics[width=\columnwidth]{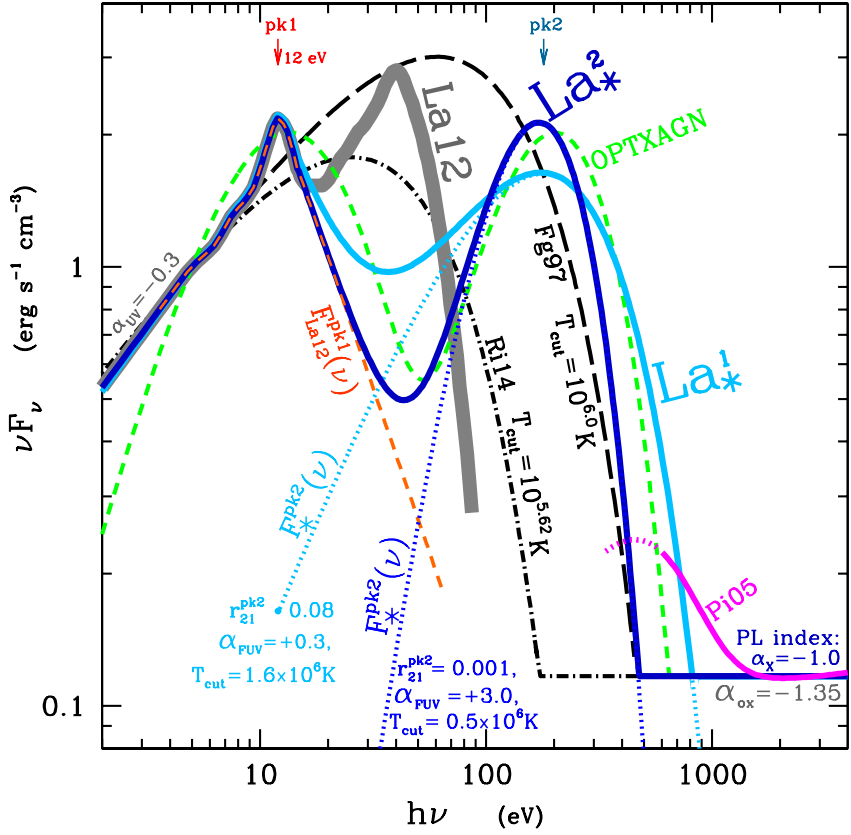}
  \caption{Ionizing spectral energy distributions described in  \S\,\ref{sec:dbump}, in $\nu F_{\nu}$ units: 1) the SED  adopted by Fg97 for their LOC calculations with $\Tcut = 10^{6.0}$\,\degk\ (long dashed line), 2) the optimized SED with $\Tcut = 10^{5.62}$\,\degk\ of {Ri14} (dot-short dash line), and 3) the double-bump reprocessed distribution of {La12} (thick grey line), and 4) two modified versions \Laasto\ (cyan) and  \Laastd\ (blue) of the La12 reprocessed distribution. The \Laasto\  (cyan) and  \Laastd\ (blue) SEDs were obtained by summing up the truncated La12 SED, $\Fpko(\nu)$ (red dashed line), to both cyan and blue \textit{dotted lines} $\Fpkt(\nu)$ distributions.  The light-green dashed line corresponds to an accretion disk model  including  optically thick Compton emission by a warm plasma. It was calculated using the OPTXAGN model in XSPEC \citep[][see \S\,\ref{sec:xspec}]{Dn12}.  The above SEDs were renormalized to $\nu F_{\nu}$ of unity at 5\,\ev\ (2480\,\AA). In the X-ray domain, a power-law of index $-1.0$ was assumed with an \aox\ of $-1.35$. The magenta line represents the average of the soft X-ray excess measurements inferred by Pi05, assuming  $\aox=-1.35$ with respect to the La12 distribution.
  }
    \label{fig:fig1}
\end{figure}

\section{A double bump ionizing energy distribution}
\label{sec:dbump}

\subsection{Standard ionizing SEDs}
\label{sec:stdsed}

The ionizing radiation from the nucleus is expected to originate from thermal emission by gas accreting onto a supermassive black hole. Although thermal in nature, the energy distribution is broader than a blackbody since the continuum emission is considered to take place from an extended disk that covers a wide temperature range. In their photoionization models, \citet[hereafter Fg97]{Fg97} and \citet[hereafter Ri14]{Ri14} assumed a SED where the dominant ionizing continuum corresponds to a thermal distribution of the form  
\begin{equation}
\label{disk_eqn}
F_{\nu} \propto \nu^{\auv} \exp(-h\nu/k\Tcut) 
\end{equation} \label{pkone_eqn}
\noindent where \Tcut\ is the UV temperature cut-off and \auv\ the low-energy slope of the `big bump', which is typically assumed to be $\auv=-0.3$.  This thermal component dominates the ionizing continuum up to the X-ray domain where a power-law of index $-1.0$ takes over. The values for \Tcut\ adopted by Fg97 and Ri14 are $10^{6.0}$ and $10^{5.62}$\,\degk, respectively. Both distributions are shown in Fig.\,\ref{fig:fig1}.
In both cases we have assumed a standard X-ray-to-optical spectral index\footnote{The X-ray spectral index is defined as  $\aox = 0.3838 \times \log(f_{2\rm kev}/f_{2500{\rm A}}$), where $f_{\rm{2keV}}$ and $f_{2500{\rm A}}$ are the fluxes at rest-frame 2\,\kev\ and 2500\AA, respectively.}   \aox\ of $-1.35$. 

\subsection{The proposition of a double-peaked SED}
\label{sec:stdori}

To address the temperature problem, \citet[hereafter La12]{La12} explored the possibility that a population of internally cold very thick ($N_{\rm H}>10^{24}\,$\cms) dense clouds ($n \sim 10^{12}\,$\cmc) covers the accretion disk at a radius of $\sim 35\,R_s$ from the black-hole, where  $R_s$ is the Schwarzschild radius. The cloud's high velocity turbulent motions blur its line emission as well as reflect the disk emission, resulting in a double-peaked SED superposed to the reflected SED. The first peak at $\sim 1100$\AA\ represents the clouds reprocessed radiation while the second corresponds to the disk radiation reflected by the clouds. The resulting SED is represented by the thick light-gray continuous line in Fig.\,\ref{fig:fig1}. The main  advantage of this distribution is its ability to account for the `universal' knee observed at 12\,\ev\ in quasars. The assumed position of the second peak at 40\,\ev\ would however need to be shifted to much higher energies in order to significantly increase the photoheating efficiency and subsequently reproduce the observed \Roiii\ ratio.  This possibility, which is explored in the current work,  might imply adjustments of the `reprocessing model' since the  turbulent clouds, hypothesised by La12, would likely need to extend to much smaller radii than the assumed value of 35\,$R_s$. Alternatively, the hotter inner component of the accretion disk might progressively become uncovered at smaller disk radii. We note that similar double-peaked SEDs would arise if the primary disk emission was further Compton up-scattered to higher energies owing to the presence of an optically thick warm plasma in addition to the hot thin corona responsible for the hard X-rays \citep{Dn12}. We will further discuss this possibility in \S\,\ref{sec:disc}.

\subsection{Components of our modified double-peak  \Laasto\ SED}
\label{sec:compone}

After experimenting with different shapes and positions for the second peak, it was found that the presence of a deep valley at $\simeq 35$\,\ev\ can result in an increase of the plasma temperature (\textit{i.e.} higher \Roiii\ ratio). To explore double-peak SEDs, we proceeded as follows. First, we extracted a digitized version of the published La12 SED. To eliminate the 40\,\ev\ peak we extrapolated the declining segment of the first peak. The resulting distribution is represented by the red dashed line labelled $\Fpko(\nu)$ in Fig.\,\ref{fig:fig1}. For the second peak, $\Fpkt(\nu)$, we adopted the formula, $\nu^{\auv} \exp(-h\nu/k\Tcut)$ (i.e. Eq.\,\ref{disk_eqn}). All the double-bump SEDs which we explored were obtained by simply summing both distributions:
\begin{equation}
\label{bump_eqn}
   F_{\nu} = \Fpko(\nu) + R \, \; \rpk  \, \nu^{\afuv} \exp{(-h\nu/k\Tcut)}
\end{equation} \label{pktwo_eqn}
\noindent where $R=\Fpko(\nu_{pk1})/\Fpkt(\nu_{pk1})$ is the renormalization factor which we define at $h \nu_{pk1}= 12\,$\ev, the energy where the first peak reaches its maximum in $\nu F_{\nu}$.  The position and width of the second peak depends on both parameters \auv\ and \Tcut\ while its intensity is set by the parameter \rpk. The main benefit of the second peak is to increase the local heating rate due to \hep\ photoionization (c.f. \S\,\ref{sec:heat}).

After comparing the plasma temperatures reached when different combinations of the parameters \Tcut, \afuv\ and \rpk\ are considered, we concluded that the optimal position for the second peak is $\approx 200\,$\ev. Moving it to higher values was not an option as it generated an excessive flux in the soft X-rays that is not observed in Type II AGN.  

Our first version for the optimal SED, labelled \Laasto, is shown in Fig.\,\ref{fig:fig1} (cyan solid line). It assumes an index $\afuv=+0.3$ as in Ri14 and Fg97, which corresponds essentially to the index of the standard Shakura-Sunyaev accretion disk model \citep{SS73,Pr81,Ch19} of $\afuv = 1/3$. The value derived for the parameter  \Tcut\ is $1.6 \, 10^6\,$\degk\ and the optimal value for the scaling factor is $\rpk = 0.08$. Increasing \rpk\ further would require a reduction in \Tcut\, otherwise, the resulting SED would extend too far into the soft X-rays.

\subsection{An alternative double-peak SED: \Laastd\ }
\label{sec:comptwo}

The \Laasto\ SED drops off around 800\,\ev\ (Fig.\,\ref{fig:fig1}). It is important to ensure that the predicted flux beyond 500\,\ev\ is not exceeding the soft X-rays measurements. While some AGN show extreme emission in the so-called X-ray soft excess up to 1--2\,\kev, others do not. One solution would be to adopt larger values for the parameter \afuv. To illustrate this, our second version of the double-bump SED, labelled \Laastd, uses a much larger \afuv\ of $+3$. In this case the optimal value for the parameter \Tcut\ has to be as low as  $0.5 \, 10^{6}\,$\degk\ in order that the second peak takes place at essentially the same energy as in the \Laasto\ SED. Because the peak profile is much narrower (dotted blue line), the scaling parameter \rpk\ turns up much smaller, at 0.001.

Note that if we had assumed the Planck equation, as in La12, for the second peak instead of Eq.\,1, the favored position of the second peak near 200\,\ev\ would correspond to a blackbody temperature of $\Tbb \simeq 500\,000\,$\degk, that is, five times higher\footnote{Corresponding to a peak shift from 40\,\ev\ to 200\,\ev.} than the $\Tbb \simeq 100\,000\,$\degk\ temperature proposed by La12.

\subsection{The unaccounted soft X-ray excess below 2\,\kev}
\label{sec:compsoft}

There are few competing processes that have been proposed to explain the physical mechanism responsible for the so called X-ray soft excess.
The most popular ones include a dual-coronal system \citep[e.g.][]{Dn12} or relativistic blurred reflection \citep[e.g.][]{RF05}.
For illustrative purposes, we show in Fig.\,\ref{fig:fig1} the ``average soft excess" component observed with XMM-Newton (magenta line) by \citet[][hereafter Pi05]{Pi05}. It corresponds to the best-fit average of 13  quasars with $z < 0.4$ using the parameters from Table\,5 of Pi05, as described in \citet{HC07}. This component was re-scaled so as to reproduce an \aox\ of $-1.35$ with respect to the La12 SED. The dotted section below 600\,\ev\ is  speculative as it is not reliably constrained by X-ray measurements. 

The soft excess varies strongly among different individual objects and its nature might be completely different than the emission in the extreme UV postulated here. On the other hand, the comptonization of disk photons by a  warm plasma can explain the presence of the soft-excess  \citep{Dn12} and at the same time produce double-peaked SEDs with the second peak near 200\,\ev\ (see Section \ref{sec:stdori}). The same might be true for blurred reflection/emission, as relativistic line emission has been used to 1) model the soft excess in a successful way \citep[e.g.][]{RF05},  and 2)  produce specific ionizing SEDs that result into two emission bumps in the extreme UV (La12, see \S\,\ref{sec:stdori}).  While tantalizing, exploring these possibilities is beyond the scope of this paper. The important point is that the proposed SEDs in this Paper are consistent with the soft excess observed in quasars.

\section{Observed \Roiii\ ratios among AGN }
\label{sec:sample}

\subsection{Seyfert\,2 samples}
\label{sec:sytwo}

In order to compare our calculations with observed ratios among Type\,II AGN, we adopt the three samples used by BVM, which are represented in Fig.\,\ref{fig:fig2} by black open symbols. They correspond to the following dereddened\footnote{All reddening corrections were carried out by the referred authors.} measurements:
\begin{enumerate} 
\item the seven Seyfert\,2's from Kos78. The average ratios is $\Roiii = 0.0168$ (i.e. $10^{-1.77}$), which is represented by a large black disk whose radius of 0.088\,dex corresponds to the RMS \Roiii\ dispersion. A unique characteristic of this sample is the availability of measurements from the weak \ariv\ $\lambda\lambda$4711,40\AA\ doublet ratio, which can be used as a direct density indicator of the plasma responsible for the high excitation lines.
\item the average of four Seyfert\,2 measurements (IC\,5063, NGC\,7212, NGC\,3281 and NGC\,1386) observed by \citet[hereafter Be06b]{Be06b}. It is represented by  a small black circle corresponding to a mean \Roiii\ of 0.0188. The pseudo error bars represent an RMS dispersion of 0.042\,dex. 
\item the high excitation Seyfert\,2 subset a41 from Ri14 (open diamond), with $\Roiii=0.0155$, representing the high ionization end of the sequence of reconstructed spectra of {Ri14}, which were extracted from a sample of 379 AGN. 
\end{enumerate}

\subsection{Spatially resolved ENLR}
\label{sec:enlr}

We superpose in Fig.\,\ref{fig:fig2} the ratios observed from the \textit{spatially resolved} emission component of AGN, the so-called ENLR, which consists of \textit{off-nuclear} line emission from plasma with  densities typically $<10^3\,$\cmc\ \citep[e.g.][]{Ta94,Be06a,Be06b}. The selected measurements are represented by the filled dark-green symbols, which stand for the following four samples: 1) the average (small filled dot) of two Seyfert\,2's and two NLRGs from \citep[hereafter BWS]{BWS}, 2) the long-slit observations of the Seyfert\,2 IC\,5063 by Be06b (pentagon), 3) the average of seven spatially resolved optical filaments from the radio-galaxy Centaurus\,A  (filled square) from \citet[hereafter Mo91]{Mo91}, and 4) the 8\,kpc {\it distant cloud} from radiogalaxy Pks\,2152$-$699 observed by \citet[hereafter Ta87]{Ta87} (large dot). Pseudo-error bars denote each sample RMS dispersion.

\subsection{Quasar sample}
\label{sec:quasar}

For illustrative purposes, we overlay in Fig.\,\ref{fig:fig2} the measurements of the NLR ratios (open grey squares) from 30 quasars of redshifts $z < 0.5$, which were studied by {BL05}. The \Roiii\ ratios are found to extend from 0.01 up to 0.2, providing clear  evidence that collisional deexcitation takes place within the NLR of Type\,I objects.

\subsection{Dataset comparison}
\label{sec:dataset}

The detailed study of BVM of the Seyfert\,2 sample from Kos78 rely on the measured density sensitive \ariv\ $\lambda\lambda$4711,40\AA\ doublet. The authors found no evidence that significant collisional deexcitation was affecting the observed \Roiii\ ratios, even after considering a power-law distribution of the densities in their plasma calculations of \Roiii\ and \ariv\ ratios. Furthermore, both Seyfert\,2's and ENLR measurements occupy a similar position in Fig.\,\ref{fig:fig2}, which is likely a consequence of LDR since detailed studies of ENLR spectra are consistent with plasma densities $\ll 10^4\,$\cmc. By contrast, quasar NLR measurements of BL05 span a wide range in \Roiii, with the lowest ratios lying close to the values seen in Seyfert\,2's and in the ENLR plasma. This dichotomy between Type\,I and II objects is likely the manifestation of the observer's perspective on the NLR as a consequence of the \textit{unified AGN geometry} whereby the densest NLR components become progressively obscured in Type\,II objects due to the observer's lateral perspective on the ionizing cone. A graphical description of such geometry is illustrated by Fig.\,2 of \citet{Be06c}.

\section{Photoionization calculations}
\label{sec:phot}

Our photoionization calculations were carried using the version \ifla\ of the code \MAP\ \citep{BM12}. Recent updates are described in Appendix\,\ref{sec:ap-A}. We compare below our models with the observed \Roiii\ as well as the \heii/\hb\ \heiirb\ ratios, assuming different ionizing continua.

\subsection{Dust-free plasma with abundances above solar}
\label{sec:dustfree}

In this Paper, we will only consider the case of a dustfree plasma. Insofar as plasma metallicities, it is generally accepted that gas abundances of galactic nuclei are significantly above solar values. The metallicities we adopt below correspond to 2.5\,\zsol, a value within the range expected for galactic nuclei of spiral galaxies as suggested by the \citet{Do14} landmark study of the Seyfert\,2  NGC\,5427 using  the Wide Field Spectrograph \citep[WiFeS:][]{Do10}. The authors determined the ISM oxygen abundances from 38 \hii\ regions spread between 2 and 13\,kpc from the nucleus. Using their inferred metallicities, they subsequently modelled the line ratios of over 100 `composite' ENLR-\hii\ region emission line spaxels as well as the line ratios from the central NLR. Their highest  oxygen abundance reaches 3\,\zsol\  (i.e. $\abunoh = 9.16$). Such a high value is shared by other observational and theoretical studies that confirm the high metallicities of Seyfert nuclei \citep{SP90,Na02,Ba08}. Our selected abundance set is twice the solar reference set of \citet{AG06}, i.e. with ${\rm O/H} = 9.8\times 10^{-4}$, except for C/H and N/H which reach four times the solar values owing to secondary enrichment.


\begin{figure}[!t]
\includegraphics[width=\columnwidth]{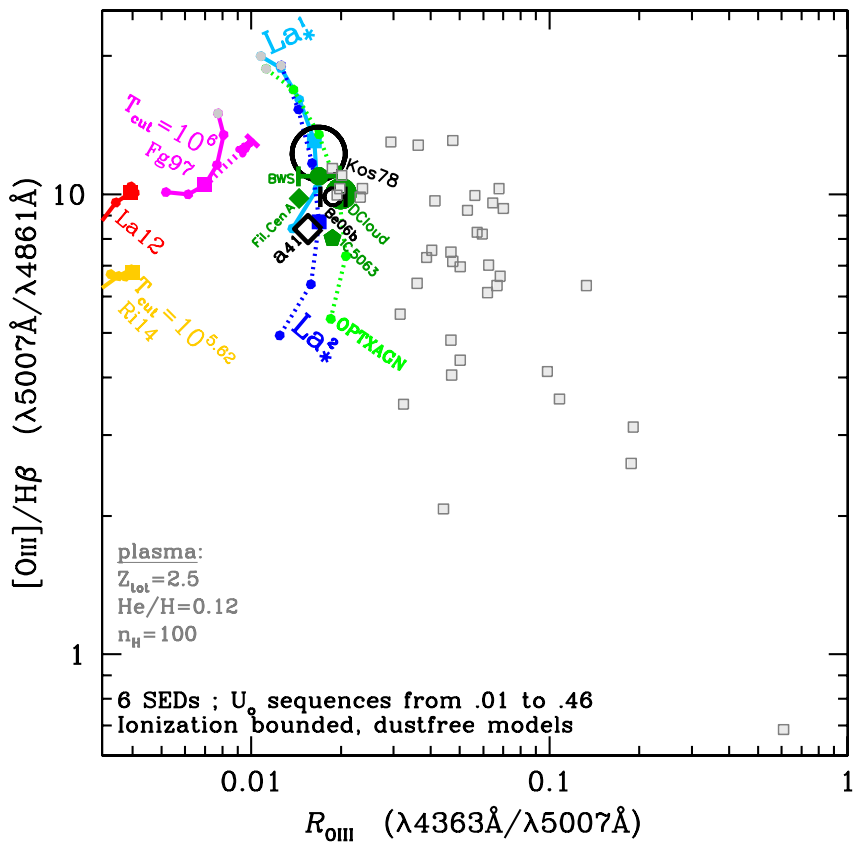}
\caption{Dereddened NLR ratios of \oiii/\hb\  vs. \Roiii.  {\it  Dataset}: line ratios from three samples of Type\,II AGN, all represented by open \textit{black} symbols and consisting of: 1) the average of seven Seyfert\,2's from Kos78 (large circle), 2) the average of four Seyfert\,2's from Be06b (small circle), 3) the high excitation Seyfert\,2 subset a41 from {Ri14} (diamond). ENLR measurements are all represented by dark-green filled symbols consisting of: 1) the average of two Seyfert\,2's and two NLRGs (small dot) from BWS, 2) the  Seyfert\,2 IC\,5063 long slit spectrum of Be06b (pentagon), 3) the average of seven spatially resolved optical filaments of the radio-galaxy Centaurus\,A  (green square) from Mo91, 4) the 8\,kpc {\it distant cloud} from radiogalaxy Pks\,2152$-$699 by Ta87 (large green dot). Type\,I AGN ratios are overlaid consisting of 30 quasars studied by {BL05} (open grey squares). {\it Models}:  Five sequences of photoionization models are overlaid along which  \uout\ increases from 0.01 (grey dot) to 0.46 in steps of 0.33\,dex, assuming a constant plasma density of $\nout=10^2$\,\cmc. A square identifies models with $\uout=0.1$. The SEDs were borrowed from Ri14 (yellow), La12 (red) and Fg97 (magenta). The dotted magenta arrow shows the effect caused by adopting a reduced abundance of 1.4\,\zsol. Calculations using the two double-peaked \Laasto\  and \Laastd\ SEDs are coded in cyan and blue colors, respectively, while those that assume comptonization of the accretion disk photons are coded in light-green.
} 
\label{fig:fig2}
\end{figure}

We can expect the enriched metallicities of galactic nuclei to be accompanied by an increase in He abundance. We followed a suggestion from David Nicholls (private communication, ANU) of extrapolating to higher abundances the metallicity scaling formulas that \citet{Ni17} derived from local\,B stars abundance determinations. At the adopted O/H ratio, the proposed scaling formula described by Eq.\,A1 in Appendix\,\ref{sec:ap-A} implies a value of He/H$\,=0.12$, which is higher than the solar ratio of 0.103 adopted by Ri14. The effect on the equilibrium temperature, however, is relatively small as the calculated \Roiii\ ratios are found to increase by only 0.06\,dex 
whether one assumes the Fg97, Ri14 or La12 SED.

\subsection{Characterization of the temperature problem}
\label{sec:tprob}

The difficulty in reproducing the observed \Roiii\ ratio is illustrated by the three ionization parameter sequences shown in Fig.\,\ref{fig:fig2} that fall on the extreme left of the diagram.  Two of the SEDs were borrowed from the standard NLR models of Ri14 (yellow) and Fg97 (magenta) while the third corresponds to the double-peaked SED from La12 (in red). 
Along each sequence, the ionization parameter\footnote{\label{noteu}$\uout = \frac{\phi_0}{c \nout}$, where $\phi_0$ is the ionizing photon flux impinging on the photoionized slab, \nout\  the hydrogen density at the {\it face} of the cloud and $c$ the speed of light.}, \uout,  increases in steps of 0.33\,dex, from 0.01 (light gray dot) up to 0.46. These sequences do not reach the  \Roiii\ domain occupied by our sample of Seyfert\,2's, with some models falling outside the plot boundaries.

\subsection{Calculations with the \Laasto\ and \Laastd\ SEDs}
\label{sec:lanew}

The procedure followed to define the double-peaked \Laasto\ and \Laastd\ energy distributions (Fig.\,\ref{fig:fig2}) have been described  in \S\,\ref{sec:compone} and \ref{sec:comptwo}. Photoionization calculations using either SED are successful in reproducing the \Roiii\ ratios from the Seyfert\,2 Kos78 sample (black circle), as shown by the solid cyan and dotted blue lines in Fig.\,\ref{fig:fig2}, which represent ionization parameter sequences with \uout\  increasing in steps of 0.33\,dex, from 0.01 (the light-gray dot) up to 0.46, assuming a constant plasma density of $\nout=10^2$\,\cmc. A square identifies models with $\uout=0.1$. Our models suggest that either of the double-bump SEDs has the potential of resolving the temperature discrepancy encountered with conventional ionizing distributions. 

We would qualify the two \Laasto\ and \Laastd\ SEDs as representing two extreme cases with respect to the parameter \afuv.
Ionizing continua that assumed intermediate values, in the range $0.3 < \afuv\ < 3$, would be equally successful in reproducing the observed \Roiii\ ratios provided the parameters  \Tcut\ and \rpk\ were properly adjusted to maintain the second peak centered at 200\,eV\ and at an intermediate height between \Laasto\ and \Laastd.


\begin{figure}[!ht]
\includegraphics[width=\columnwidth]{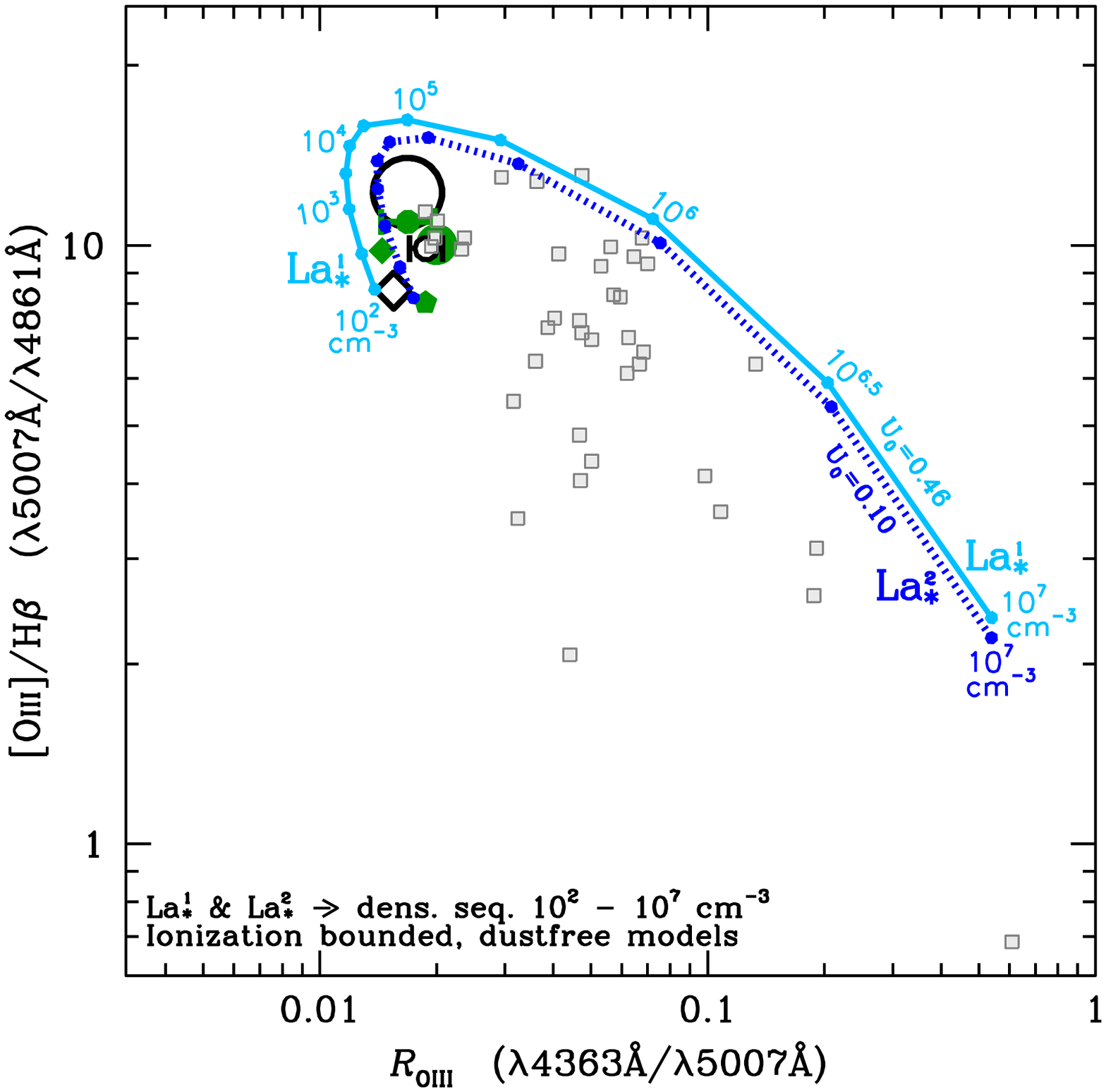}
\caption{Dereddened  \oiii/\hb\ (\oiiirb) and \heii/\hb\ (\heiirb) line ratios. The observational datasets of \S\,\ref{sec:sample} are represented by the same symbols as in Fig.\,\ref{fig:fig2}.  Are overlaid two sequences of models along which \nout\ increases from $10^2$ to $10^{7}$\,\cmc\  in steps of 0.5\,dex. The ionization parameter is $\uout=0.46$ for the \Laasto\ sequence (cyan) and $\uout=0.10$ for the  \Laastd\ sequence (blue).    
 }
\label{fig:fig3}
\end{figure}

In order to compare our models with Type\,I AGN (open grey  squares), we calculated density sequences along which the density increases in steps of 0.5\,dex, from $\nout = 100$ to $10^7$\,\cmc. These calculations are shown in Fig.\,\ref{fig:fig3}. For each sequence, we selected the \uout\ value that made the models cover the upper envelope of the quasar \oiii/\hb\ ratios, which are $\uout=0.46$ and 0.10 for the \Laasto\ and \Laastd\ SEDs, respectively. The vertical dispersion in the observed \oiii/\hb\ ratios is noteworthy. The simplest interpretation might be the need of considering a distribution of cloud densities, as favored by the dual-density models of BL05. Alternatively, in the case of the Ri14 LOC models, the observed dispersion suggests that the density power-law index $\beta$ might take on different values.

\subsection{The \heii/\hb\ diagnostic ratio}
\label{sec:heii}

When comparing different ionizing distributions, an important  ratio to consider is \heii/\hb\ \heiirb\  since, as pointed out by Ri14, the latter is sensitive to the hardness of the ionizing continuum. Fig.\,\ref{fig:fig4} illustrates the behaviour of the dereddened \oiii/\hb\  vs. \heii/\hb\ ratios assuming either the \Laasto\ or the \Laastd\ ionizing continua. They reproduce reasonably well the \heii/\hb\ ratio observed among the Seyfert\,2's of Kos78 and Be06b. Also overlaid is the \gam\ sequence from the LOC calculations (yellow dashed line) of Ri14, assuming a density weighting parameter $\beta$ of $-1.4$, the value favored by the authors when modeling their four AGN subsets\footnote{Each subset represents a composite emission line spectrum assembled from a sample of Seyfert\,2 spectra of the SLOAN database. They form an ionisation sequence in a BPT diagram that covers the locus of AGN, as described by the authors. The high excitation a41 subset is best reproduced assuming a weighting parameter \gam\ value of $\simeq -0.75$.}.


\begin{figure}[!ht]
  \includegraphics[width=\columnwidth]{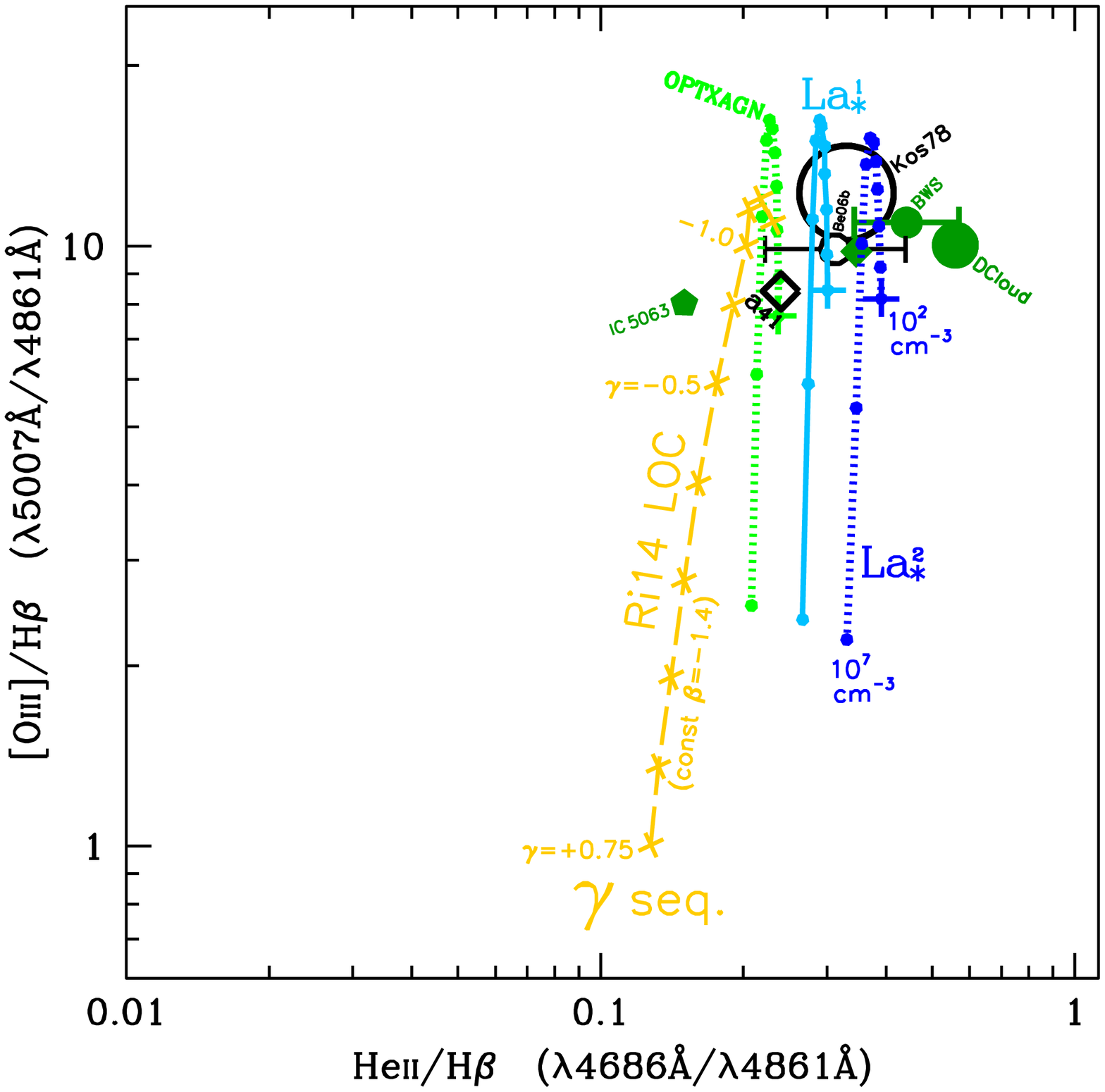}
 \caption{Dereddened  \oiii/\hb\ (\oiiirb) and \heii/\hb\ (\heiirb) line ratios. The observational datasets of \S\,\ref{sec:sample} are represented by the same symbols as in Fig.\,\ref{fig:fig2}.  Both density sequences from previous Fig.\,\ref{fig:fig3},  using the \Laasto\ (cyan) and \Laastd\  (blue) SEDs, are overlaid.  The light-green dotted line represents a density sequence assuming the accretion disk SED derived using the OPTXAGN routine. In each sequence, a cross identifies the lowest density model with $\nout=100\,$\cmc. The yellow dashed line represents the dustfree LOC model sequence from Ri14 with $\beta = -1.4$, along which the radial weighting parameter \gam\ varies from +0.75 to $-2.0$ in steps of $-0.25$.
 }
\label{fig:fig4}
\end{figure}

If we compare the Type\,II samples (black open symbols) with the spatially resolved ENLR (dark-green symbols), we notice  a wider dispersion among the \heii/\hb\ ratios than for the \Roiii\ ratios of Fig.\,\ref{fig:fig3}, which is surprising given the fact that the \heii/\hb\ ratio depends little on density or temperature. This could be an indication that the emitting plasma  in some cases is not fully ionization-bounded, as proposed by {BWS}.

\section{Discussion}
\label{sec:disc}

\subsection{Plasma heating from \hep\ photoionization}
\label{sec:heat}

The presence  of high excitation lines among NLR spectra such as \heii, \ciii, \civ, \neiii\  indicates a hard ionizing continuum. As a consequence of the SED hardness, the heating rate as well as the resulting equilibrium temperatures are higher than in \hii\ regions due to the higher energies of the ejected photoelectrons and to the significant contribution of \hep\ photoionization to the total heating rate, at least within the front layers of the exposed nebulae. The fraction of ionizing photons with  energies above 54.4\,\ev\ is 24\,\% and 26\,\% for the  \Laastd\ and \Laasto\ SEDs, respectively, and 23\,\% for the Fg97 distribution\footnote{The fraction is  19\,\% for the OPTXAGN SED.}. These values are quite similar, but when a dip takes place below 50\,\ev, as in the  \Laastd\ and \Laasto\ SEDs (see Fig.\,\ref{fig:fig1}), the peak of the distribution in $\nu F_{\nu}$ shifts to higher energies ($\simeq 200\,$\ev). Consequently, the plasma heating rate\footnote{The heating rate is the result of the thermalisation of the photoelectrons ejected from the \hi, \hep\ and, to a lesser extent, \hen\ species.} rises above the rate obtained with the Fg97 SED, essentially as a result of the increase in the mean energy of the photoelectrons ejected from ionization of \hep. 

\subsection{The shape of the EUV dip near 40\,\ev }
\label{sec:dip}

Being able to reproduce the observed \Roiii\ measurements of Seyfert\,2's by assuming the above double-peaked \Laasto\  and \Laastd\ SEDs does not prove it is the right solution to the temperature problem, but it is a possibility worth exploring further. One advantage of the proposed distributions is that, unlike the conventional thermal SEDs of Ri14 or Fg97, they incorporate the `universal' knee  observed near 12\,\ev\ in high redshift quasars, which \citet{ZK97} and \citet{Te02} studied using the  \textit{Hubble Space Telescope} archival database.  Using the Far Ultraviolet Spectroscopic Explorer database, \citet{Sc04} find an average index \alnu\ of  $-0.56$ in AGN of redshifts $<0.33$, which is significantly flatter than the average of $\simeq -1.76$ found at redshifts $z \ga 2$  \citep{Te02}. How steeply the flux declines beyond 12\,ev\ is rather uncertain and could also depend on the AGN luminosity \citep{Sc04}.

\subsection{Comptonized Accretion disk models}
\label{sec:xspec}

The pertinence of a second peak to describe the harder UV component is  provided by the work of \citet{Dn12} who built a self-consistent accretion model where the primary emission from the disc is partly comptonized by an optically thick warm plasma, forming the EUV. This plasma that according to \citet{Dn12} might itself be part of the disk would exist in addition to the optically thin hot corona above the disc responsible for producing the hard X-rays. 

Using the OPTXAGN routine in XSPEC\footnote{Using commands described in http://heasarc.gsfc.nasa.gov/xanadu/xspec/manual/node132.html \citep{Dn12,Ku18}}, we calculated an accretion disk model that allowed the second peak to occur at a similar position as that of the \Laastd\ SED. It is represented in Fig.\,\ref{fig:fig1} by the light-green dashed curve. Different sets of parameters in the OPTXAGN model can match the double peaked \Laasto\ and \Laastd\ SEDs, provided extreme accretion rates are assumed (L/L$_{Edd}\geq$1). Such accretion rates are not proper of the Type II objects discussed here, but rather of extreme Narrow Line Seyfert\,1 nuclei. Even though the OPTXAGN model was not developed to generate the double peak SEDs postulated in this work, we should note that a wide set of SEDs with different double peaks or even more extreme FUV peaks can be produced using different, less stringent, parameters.  Our aim in presenting this SED, apart from matching our \Laastd\ SED, was to exemplify how different physical processes at different accretion disk scales might result in an ionizing distribution capable of reproducing the observed \Roiii\ ratio. We plan to fully explore under which conditions (e.g. more conservative SEDs with moderate accretion rates), dual temperature comptonization disk models can achieve this.

An additional drawback is that the OPTXAGN model does not reproduce the knee observed at 12\,\ev. Expanding the range of parameters in this model might circumvent this issue. Overall, we note the striking similarity of the OPTXAGN with that of \Laastd, considering that they were built independently and with completely different scientific motivations.     
Photoionization calculations with this SED indicate that, as expected, it can reproduce the observed \Roiii\ ratio, as shown by the light-green dotted line in Fig.\,\ref{fig:fig2}. The \heii/\hb\ ratio, however, is somewhat under-predicted, as shown in Fig.\,\ref{fig:fig2}. This appears to be caused by the first peak of the OPTXAGN SED being significantly thicker than in the \Laastd\ SED.

\begin{acknowledgements}

AH acknowledges support from NVIDIA in the form of a GPU under the NVIDIA Academic Hardware Grant Program. SGM acknowledges support from the Funda\c{c}\~{a}o para a Ci\^{e}ncia e a Tecnologia (FCT) through the Fellowships PD/BD/135228/2017 (PhD::SPACE Doctoral Network PD/00040/2012), POCH/FSE (EC) and COVID/BD/152181/2021.
AH and SGM were also supported by Funda\c{c}\~{a}o para a Ci\^{e}ncia e a Tecnologia (FCT) through the research grants UIDB/04434/2020 and UIDP/04434/2020, and an FCT-CAPES Transnational Cooperation Project "Parceria Estrat\'egica em Astrof\'{\i}sica Portugal-Brasil". 

\end{acknowledgements}

\begin{appendices}
\setcounter{equation}{0}

\section{Recent updates to the code \map} 
\label{sec:ap-A} 

We incorporated the following tools in the  version \ifla\ of \MAP.

\begin{list}{}{}

\item[a)] We implemented the new routine \OSALD\, which calculates various line ratio diagnostics that can be used to infer the temperature and/or density cut-off using observed line ratios. It assumes an isothermal plasma that covers a wide range of densities, up to a predefined cut-off density. The diagnostics can also be applied to line ratios not previously dereddened since \OSALD\ offers the option of dereddening the line ratios from the observed Balmer lines. It is also possible to assume a dust extinction that correlates with the plasma density, a possibility relevant to the NLR of Type\,II AGN. The routine is described in \S\,5 (and Appendix\,C) of BVM. 

\item[b)] Based on the work of \citet{PPB91}, the recombination rates from \npp, \oppp\ and \opp\ to the corresponding metastable levels \nso\ and \ndd\ of \nii\ and \oiii\ and levels \dpx\ and \ddx\ of \oii, respectively have been incorporated in the calculation of the corresponding emission line intensities. In the case of the \nso\ level of \oiii, we added the missing contribution from dielectronic recombination (Christophe Morisset, private communication). As for the \spp\ and \sppp\ ions, we estimated their recombination rates to metastable levels by extrapolation from the \opp\ and \oppp\ ions as follows: we assumed that the fraction, \ram, of the total recombination rate ($\alpha_{\rm SII}^{rec}$ or $\alpha_{\rm SIII}^{rec}$), which populates metastable levels of Sulphur is the same fraction as found for Oxygen. For instance, for a 10\,000\,K plasma this fraction, \ramoiii, in the case of level \nso\ of  \oiiix\ (responsible for the emission of the \oiiitw\ line) is $2.2$\% of $\alpha_{\rm OIV}^{rec}$. 

\item[c)] An option is offered to scale the He abundance in function of the oxygen abundance, in accordance to the Equation: 
\begin{equation}
   {\rm He/H}  =  0.06623 + 0.0315 \, (\frac{{\rm O/H}}{5.75 \times 10^{-4}})
\end{equation} \label{he_eqn}

\noindent which follows a suggestion from David Nicholls (private communication, ANU). It differs from Eq.\,4 of \citet{Ni17} as it behaves linearly down to primordial abundances. If we assume the O/H ratio given by the solar abundance set of \citet{AG06}, the He/H ratio\footnote{The solar He/H ratio adopted by Ri14 with CLOUDY is 0.103.} derived is 0.093. Beyond solar metallicities, it remains an open question to what extent the He/H ratio of the ISM from nuclear regions exceeds the solar neighborhood value. Eq.\,A1 is intended as an exploratory tool to study the impact of using above solar He/H ratios when modelling the emission plasma from a metallicity enriched interstellar medium. The value of ${\rm He/H} = 0.12$ referred to in Sect.\,\ref{sec:phot} is based on our adopted abundance set of $\ztot=2.5$, which has ${\rm O/H} = 9.8\times 10^{-4}$ as defined in Sect.\,\ref{sec:phot}. The inferred He/H ratio differs slightly from the value of 0.107 obtained using Eq.\,4 of \citet{Ni17}.

\end{list}

\end{appendices}

\bibliography{Seyfert_2_SED}
 
\end{document}


\bibliography{biblio}

\end{document}
